\documentclass[conference]{IEEEtran}
\IEEEoverridecommandlockouts
\usepackage{cite}
\usepackage{amsmath,amssymb,amsfonts}
\usepackage{algorithmic}
\usepackage{graphicx}
\usepackage{textcomp}
\usepackage{xcolor}
\usepackage{epsfig}
\usepackage{balance}
\usepackage[printonlyused]{acronym}
\usepackage{graphicx}
\usepackage{grffile}
\usepackage{flushend}

\acrodef{bs}[BS]{base station}
\acrodef{bss}[BSs]{base stations}
\acrodef{pl}[PL]{path loss}
\acrodef{ple}[PLE]{path loss exponent}
\acrodef{pdp}[PDP]{power delay profile}
\acrodef{cir}[CIR]{channel impulse response}
\acrodef{mpc}[MPC]{multi-path component}
\acrodef{mpcs}[MPCs]{multi-path components}

\def\BibTeX{{\rm B\kern-.05em{\sc i\kern-.025em b}\kern-.08em
    T\kern-.1667em\lower.7ex\hbox{E}\kern-.125emX}}
    
\title{\textsc Hybrid TRP–UE Sensing for\\ Enhanced Target Localization}

\author{\IEEEauthorblockN{Necati Kagan Erkek\IEEEauthorrefmark{1}\IEEEauthorrefmark{2}, Marco Di Renzo\IEEEauthorrefmark{1}\IEEEauthorrefmark{3}, Arman Shojaeifard\IEEEauthorrefmark{2} \\ Yasser Mestrah\IEEEauthorrefmark{2}, Remun Koirala\IEEEauthorrefmark{2}, Mohammad Heggo\IEEEauthorrefmark{2}, Kunjan Shah\IEEEauthorrefmark{2}}
\IEEEauthorblockA{\IEEEauthorrefmark{2}InterDigital, Inc., London EC3A 3DH, United Kingdom}
\IEEEauthorblockA{\IEEEauthorrefmark{1}King's College London, Centre for Telecommunications Research, London WC2R 2LS, United Kingdom }
\IEEEauthorblockA{\IEEEauthorrefmark{3}Universit\'e Paris-Saclay, CNRS, CentraleSup\'elec, Laboratoire des Signaux et Syst\`emes, 91190 Gif-sur-Yvette, France}
\IEEEauthorblockA{E-Mail: necati.erkek@kcl.ac.uk}

}

\begin{document}
\maketitle

\begin{abstract}
Integrated Sensing and Communication (ISAC) refers to the capability for the network to provide communications services whilst also being able to sense the environment in a scalable manner. One of the key functions of ISAC is the accurate localization of passive and mobile sensing targets. This paper introduces a novel hybrid TRP-UE sensing mechanism that improves network-based sensing performance. Evaluation results are provided using 3GPP-compliant ISAC channel models. The results demonstrate the significant benefit in complimenting TRP-based sensing with UE-assisted sensing in challenging propagation environments such as indoor factory. 
\end{abstract}


\section{Introduction}

Integrated Sensing and Communication (ISAC) refers to communication infrastructures that not only provide data transmission but also perform sensing tasks—such as detection, tracking, and localization—by utilizing the same spectrum, waveforms, and hardware resources \cite{10077115,10577673,ZhangJSTSP2021,Liu2022Survey,Liu2020JCR}. This multifunctional paradigm represents a fundamental shift from previous generations of wireless systems, where sensing and communication were designed and operated independently. By contrast, ISAC enables both transmission-reception-point (TRP) and user equipment (UE) to sense their surrounding environment with high spatial and temporal resolution, thereby supporting a wide range of emerging applications including environmental monitoring and advanced localization services. 

\begin{figure}[!t]
    \centering
    \includegraphics[width=0.95\linewidth]{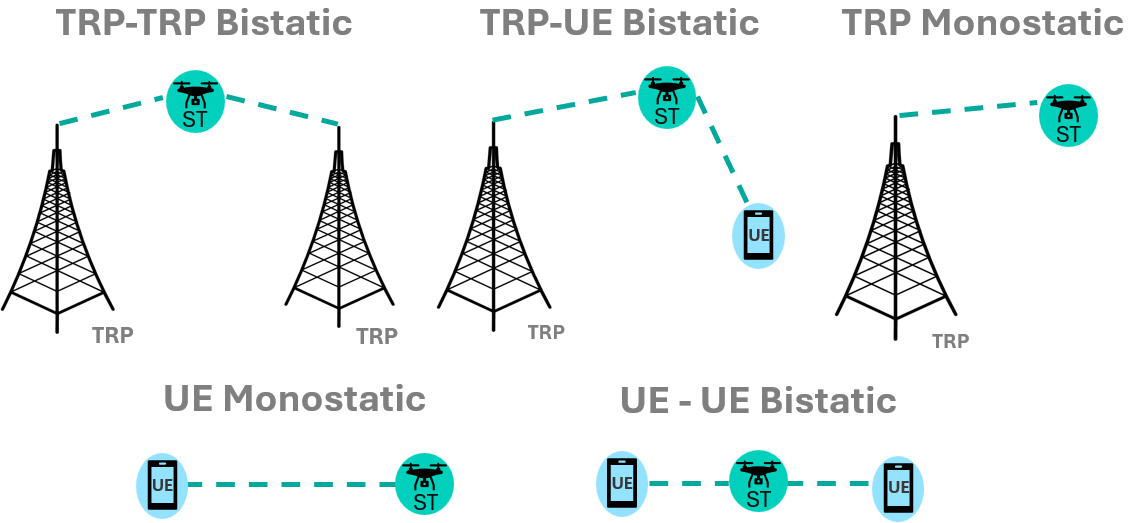}
    \caption{Illustration of different ISAC sensing modes.}
    \label{sensing_modes}
\end{figure}

Future ISAC systems may support several sensing modes, involving different combinations of TRPs and UEs \cite{ZhangVTM2021}. These sensing modes include both monostatic and bistatic modes, which are illustrated in Figure~\ref{sensing_modes}. In monostatic sensing modes, a single TRP or UE performs both the transmission and reception operations, functioning similarly to traditional radar \cite{Bistatic3GPPVTC2024}. In contrast, bistatic sensing modes distribute the transmitter and receiver across different nodes or locations. This spatial separation can provide additional geometric diversity, which is beneficial for improving target observability and reducing localization uncertainty. Moreover, combining measurements from multiple sensing modes enables the network to exploit complementary propagation paths and adapt the sensing configuration according to deployment conditions. To support these capabilities, standardization efforts are underway. For example, the 3GPP Release~19 study on ISAC has introduced channel models tailored to joint sensing and communication \cite{3gppTR22837}. Many of the envisioned ISAC use cases require accurate localization of passive targets--such as unmanned aerial vehicles (UAVs) and vehicles--based on radio-frequency (RF) signal reflections \cite{10978368,10680116,10833623,8902085}.


\begin{figure}[!ht]
    \centering
    \includegraphics[width=0.75\linewidth, height=0.6\linewidth]{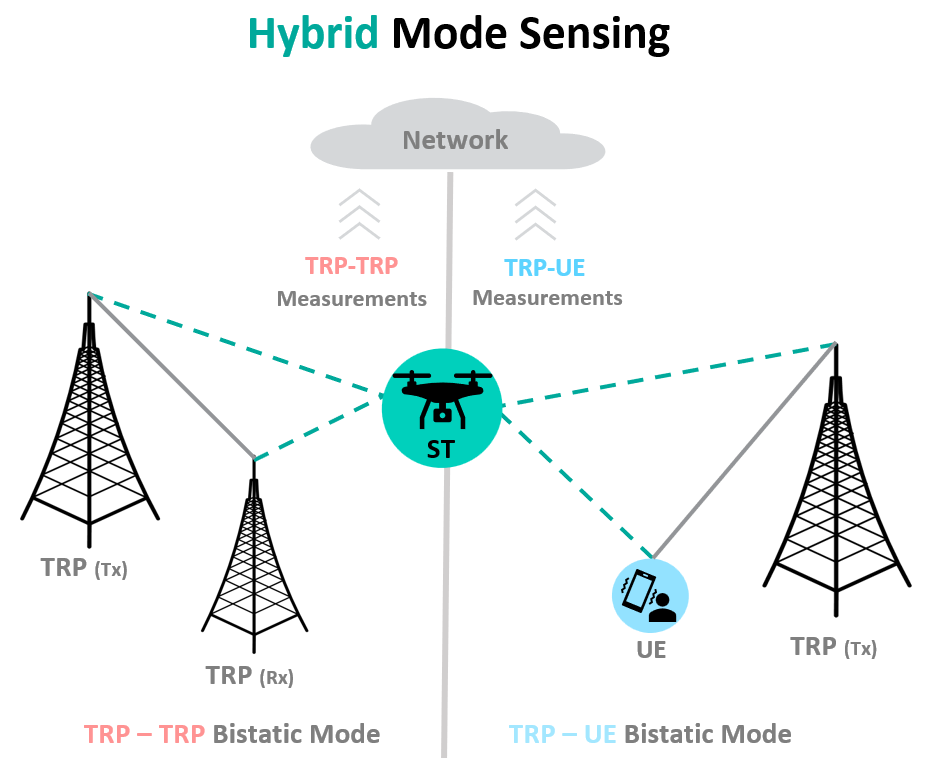}
    \caption{Hybrid sensing mode combining the measurements from TRP--TRP and TRP--UE bistatic links, to improve sensing target localization accuracy.}
    \label{Hybrid_Mode_Sensing}
\end{figure}

The work presented in this paper compares different sensing modes and introduces a novel hybrid TRP--UE sensing mechanism, in which TRP--TRP and TRP--UE bistatic measurements are fused to improve the localization quality of a sensing target. As illustrated in Fig.~\ref{Hybrid_Mode_Sensing}, the proposed data fusion approach enhances localization performance by exploiting the complementary characteristics of each sensing path. In particular, TRP--TRP sensing can provide network-side geometric diversity and stable infrastructure-based observations, while TRP--UE sensing can introduce additional spatial perspectives through user equipment acting as sensing-related nodes. By combining these heterogeneous measurements, the network can obtain a richer representation of the target environment, which can reduce localization ambiguity and improve robustness under challenging propagation conditions.

A dedicated sensing function (SF) within the network is responsible for collecting, processing, and fusing the available sensing measurements before performing target localization. The collected data may consist of delay-based measurements or joint time-angle measurements, depending on the sensing configuration selected by the network and the capabilities of the participating nodes. This flexibility allows the hybrid sensing framework to support different deployment scenarios, ranging from simpler delay-only configurations to more advanced configurations that exploit both temporal and angular information. Accordingly, the proposed hybrid sensing concept aligns with the sensing and positioning capabilities envisioned for future wireless systems, where integrated communication and sensing are expected to support more accurate, reliable, and context-aware localization services. 

\section{Signal and Noise Modelling}

Accurate localization of a sensing target depends fundamentally on the quality of received signals and the characteristics of system noise. This section presents a modeling framework for both received power and noise power to establish the signal-to-noise ratio (SNR), which is a key indicator of signal fidelity. The SNR directly influences the precision of localization techniques, as higher SNR levels enable more accurate estimation of key parameters such as angle of arrival (AoA) and time of arrival (ToA) \cite{erkek2026communication, erkek2026spaceborne, erkek2026reference, erkek2026joint, erkek2026pilot}. A clear understanding of these signal and noise interactions is therefore essential for designing robust sensing and positioning systems.

\subsection{Received Power \( P_r \)}

The received power is modeled using the Friis transmission equation with path loss \cite{10012846, erkek2026communication}:
\begin{equation}
    P_r = P_t + K - 10\alpha \log_{10}\!\left(\frac{d}{d_r}\right)
\end{equation}

Here \(P_t\) is the transmit power in dBm; \(\alpha\) is the path-loss exponent determined by the indoor factory environment; \(d\) is the distance between the User Equipment (UE) and the Transmission Reception Point (TRP); and \(d_r\) is a reference distance, typically \(1\) m. The term
\(K = 20 \log_{10}\!\bigl(\tfrac{\lambda}{4\pi d_r}\bigr)\) (in dB) collects the wavelength \(\lambda\) and the reference distance into a constant intercept for the log-distance model.

\subsection{Noise Power \( N \)}

The receiver noise power combines thermal noise and the receiver’s noise figure:
\begin{equation}
    N = N_0 \, B, \qquad N_0 = k\,T\,F 
\end{equation}

In these expressions, \(k = 1.38 \times 10^{-23}\,\mathrm{J/K}\) is Boltzmann’s constant, \(T = 290\,\mathrm{K}\) is the standard noise temperature, \(F = 10^{NF/10}\) is the linear noise factor corresponding to a noise figure \(NF\) in dB, and \(B\) is the receiver bandwidth in Hz. The resulting noise power (in watts) is converted to dBm for consistency with other link-budget terms as
\begin{equation}
    N_{\mathrm{dBm}} = 10 \log_{10}(N) + 30 
\end{equation}

\subsection{Signal-to-Noise Ratio (SNR)}

The SNR in dB is defined by the difference between received signal power and noise power: 

\begin{equation}
    \mathrm{SNR} = P_r - N_{\mathrm{dBm}}, \qquad
    \mathrm{SNR}_{\mathrm{linear}} = 10^{\mathrm{SNR}/10}
\end{equation}

 The linear SNR is used to perform Cramér–Rao bound calculations and other estimation-theoretic performance metrics that require a power ratio.

\section{Error Modelling}

\subsection{CRLB-Driven Measurement Error Model}

Imperfect sensing is modeled by perturbing ideal angles and delays with zero-mean Gaussian noise whose variances equal the corresponding Cramér–Rao lower bounds (CRLB). Using the linear SNR, \(\mathrm{SNR}_{\mathrm{linear}}\), the measured (azimuth/elevation) angle(s) and delay for each sensing link \(l\) are generated as
\begin{equation}
    \tilde{\theta}_{l} = \theta_{l} + \varepsilon_{\theta,l}, 
    \qquad \varepsilon_{\theta,l} \sim \mathcal{N}\!\left(0,\, \mathrm{CRLB}_{\theta,l}\right)
\end{equation}
\begin{equation}
    \tilde{\tau}_{l} = \tau_{l} + \varepsilon_{\tau,l}, 
    \qquad \varepsilon_{\tau,l} \sim \mathcal{N}\!\left(0,\, \mathrm{CRLB}_{\tau,l}\right)
\end{equation}
with errors independent across links and across angle/delay components. Closed-form CRLB expressions for angle- and delay-based measurements follow \cite{17564}:
\begin{equation}
    \mathrm{CRLB}_{\theta,l} = \frac{6}{m_l^{3}\,\mathrm{SNR}_{\mathrm{linear},l}}
\end{equation}
\begin{equation}
    \mathrm{CRLB}_{\tau,l} = \frac{3\,c^{2}}{\mathrm{SNR}_{\mathrm{linear},l}\,(4\pi^{2})\,B^{2}} 
\end{equation}

Here \(m_l\) denotes the number of antenna elements for link \(l\), \(c\) is the speed of light, and \(B\) is the system bandwidth. For a sensing target with true position \(\mathbf{p}\) and estimate \(\hat{\mathbf{p}}\), the localization error vector is
\begin{equation}
    \mathbf{e} = \hat{\mathbf{p}} - \mathbf{p}
\end{equation}

\subsection{LOS Probability Model}

In indoor industrial environments, the probability of a Line-of-Sight (LOS) path between a transmitter and receiver substantially affects signal strength, multipath, and system performance. For the Indoor Factory Dense Clutter (InF-DL) scenario, we adopt the ETSI TR 138~901 V17.0.0 (2022-04) model in which the LOS probability decays exponentially with the 2D separation \(d_{2D}\):
\begin{equation}
    \Pr_{\mathrm{LOS}}(d_{2D}) = \exp\!\left(-\frac{d_{2D}}{k_{\mathrm{InF\text{-}SL}}}\right)
\end{equation}

If \(r\) denotes the LOS probability at the cutoff distance \(d_{2D}=d_{\mathrm{cutter}}\), then the decay constant is
\begin{equation}
    k_{\mathrm{InF\text{-}SL}} = -\,\frac{d_{\mathrm{cutter}}}{\ln r}
\end{equation}

In this formulation, \(d_{\mathrm{cutter}}\) (in meters) marks the range beyond which LOS becomes significantly less likely, and \(r \in (0,1)\) parameterizes that likelihood at the cutoff. This probabilistic model is used in simulation to weight received power and SNR by the likelihood of LOS, thereby capturing realistic factory propagation conditions.

\section{Positioning Algorithms}
Accurate target localization requires mathematical algorithms capable of translating physical measurements (e.g., time delay, angle of arrival) into spatial coordinates. This section presents the two primary localization algorithms evaluated in this work—delay-based and time–angle-based—and explains their operation in three-dimensional (3D) space.


\subsection{3D Delay Based Positioning Algorithm}

This method estimates the spatial location of the sensing target (ST) using measured delays from four different transmitter-receiver pairs (TRPs and UEs). These delays define ellipsoids with the ST lying on their surfaces. Each ellipsoid has two foci—one at the TRP and the other at the UE—and the total path corresponds to the measured delay, as illustrated below.

\begin{figure}[!ht]{
    \centering
    \includegraphics[width=0.9\linewidth, height=0.75\linewidth]{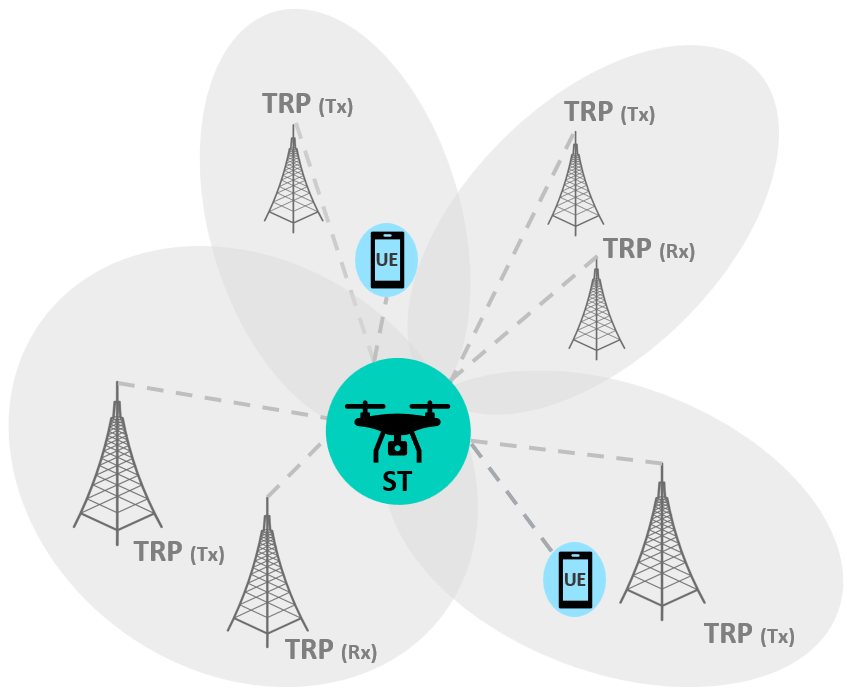}
    \caption{Four different pairs and ellipsoids with foci points.}
    \label{delay}}
\end{figure}

\subsubsection{Matrix Formulation}
For each of the four ellipsoids, we write the general ellipsoid equation in terms of $x$, $y$, and $z$. For the first ellipsoid, this takes the form:

\begin{equation}
\frac{(x - x_1)^2}{a_1^2} + \frac{(y - y_1)^2}{b_1^2} + \frac{(z - z_1)^2}{c_1^2} = 1
\end{equation}

Expanding the yields terms with respect to $x$, $y$, and $z$:

\begin{itemize}
    \item $x$-term: $-\frac{2x_1}{a_1^2} x$, $y$-term: $-\frac{2y_1}{b_1^2} y$, $z$-term: $-\frac{2z_1}{c_1^2} z$
\end{itemize}

Repeating this for each ellipsoid gives a system of equations that can be written in matrix form.

\subsubsection{Constructing Matrix $A$ and Vector $b$}

From the expanded ellipsoid equations, we can define matrix $A$ and vector $b$ as:

\begin{equation}
A = 
\begin{bmatrix}
\frac{2x_1}{a_1^2} & \frac{2y_1}{b_1^2} & \frac{2z_1}{c_1^2} \\
\frac{2x_2}{a_2^2} & \frac{2y_2}{b_2^2} & \frac{2z_2}{c_2^2} \\
\frac{2x_3}{a_3^2} & \frac{2y_3}{b_3^2} & \frac{2z_3}{c_3^2} \\
\frac{2x_4}{a_4^2} & \frac{2y_4}{b_4^2} & \frac{2z_4}{c_4^2}
\end{bmatrix}, \quad
b = 
\begin{bmatrix}
\frac{x_1^2}{a_1^2} + \frac{y_1^2}{b_1^2} + \frac{z_1^2}{c_1^2} - 1 \\
\frac{x_2^2}{a_2^2} + \frac{y_2^2}{b_2^2} + \frac{z_2^2}{c_2^2} - 1 \\
\frac{x_3^2}{a_3^2} + \frac{y_3^2}{b_3^2} + \frac{z_3^2}{c_3^2} - 1 \\
\frac{x_4^2}{a_4^2} + \frac{y_4^2}{b_4^2} + \frac{z_4^2}{c_4^2} - 1
\end{bmatrix}
\end{equation}

Each row corresponds to an ellipsoid, and the constants on the right-hand side depend on the ellipsoid parameters derived from delay measurements.

\subsubsection{Solving the Least Squares Problem}

Once matrices $A$ and $b$ are formed, we solve for the estimated location vector $v = [x \quad y \quad z]^T$ using the least squares approach:

\begin{equation}
v = (A^T A)^{-1} A^T b
\end{equation}

To ensure numerical stability, degenerate rows or columns are removed if any of the ellipsoids collapses (e.g., \(b_i=0\) or \(c_i=0\)). This approach minimizes the squared error between measured and theoretical ellipsoid surfaces, yielding the most probable 3D target position.

\subsection{3D Time-Angle Based Positioning Algorithm}

Based on the 3GPP TR 38.901 specification, in the 3D case, the Angle of Arrival (AoA) at the User Equipment (UE) or the Angle of Departure (AoD) at the TRP is decomposed into two components such as elevation angle and azimuth angle.

\begin{figure}[!ht]{
    \centering
    \includegraphics[width=0.8\linewidth]{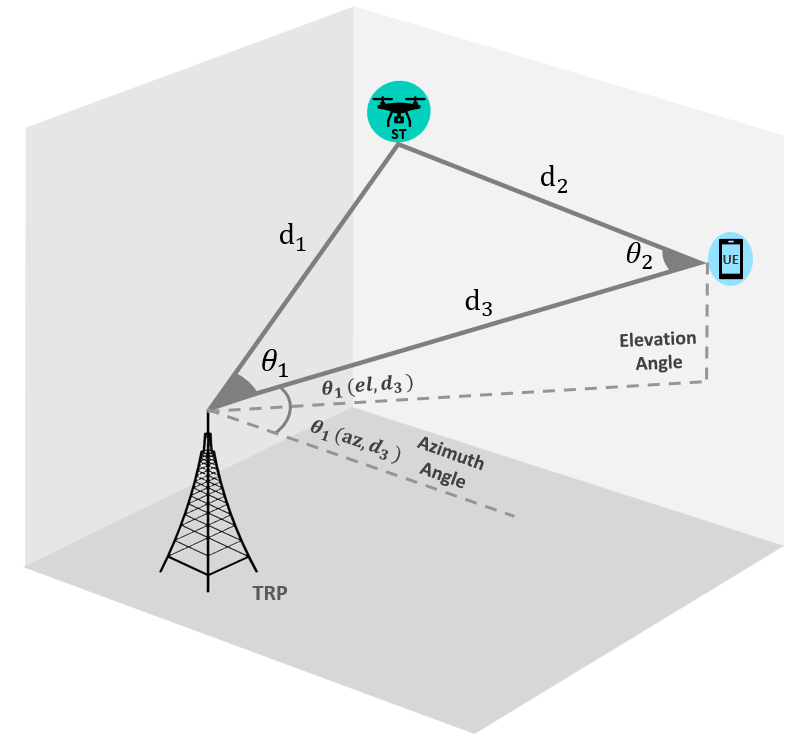}
    \caption{3D demonstration of Time-Angle based algorithm.}
    \label{angle_time}}
\end{figure}

Here, \( \theta_{\text{el}} \) is the angle between the transmitted/received beam and the positive \textit{z}-axis, and \( \theta_{\text{az}} \) is the angle between the beam and the positive \textit{x}-axis (projected on the \textit{x-y} plane). Let \( d_1 \) be the distance between TRP and the sensing target, \( d_2 \) be the distance between the UE and the sensing target, \( d_3 \) be the distance between TRP and UE, and \( d_4 \) be the total path length from TRP to the target and then to the UE, i.e., \( d_4 = d_1 + d_2 \). The AoD for the path including \( d_1 \) and the AoA for the path including  \( d_2 \) is expressed as:

\begin{equation}
(\theta^{\text{AoD}}_{\text{el}, d_1}, \theta^{\text{AoD}}_{\text{az}, d_1}), 
\quad
(\theta^{\text{AoA}}_{\text{el}, d_2}, \theta^{\text{AoA}}_{\text{az}, d_2})
\end{equation}

The intermediate directions \( A_1, A_2, A_3 \) are given by:
\begin{equation}
\begin{aligned}
A_1 &= \left\langle
\begin{matrix}
\cos(\theta^{\mathrm{AoD}}_{\mathrm{el}, d_1}) \cos(\theta^{\mathrm{AoD}}_{\mathrm{az}, d_1}),\\
\cos(\theta^{\mathrm{AoD}}_{\mathrm{el}, d_1}) \sin(\theta^{\mathrm{AoD}}_{\mathrm{az}, d_1}),\\
\sin(\theta^{\mathrm{AoD}}_{\mathrm{el}, d_1})
\end{matrix}
\right\rangle,\\[4pt]
A_2 &= \left\langle
\begin{matrix}
\cos(\theta^{\mathrm{AoA}}_{\mathrm{el}, d_2}) \cos(\theta^{\mathrm{AoA}}_{\mathrm{az}, d_2}),\\
\cos(\theta^{\mathrm{AoA}}_{\mathrm{el}, d_2}) \sin(\theta^{\mathrm{AoA}}_{\mathrm{az}, d_2}),\\
\sin(\theta^{\mathrm{AoA}}_{\mathrm{el}, d_2})
\end{matrix}
\right\rangle,\\[4pt]
A_3 &= \left\langle
\begin{matrix}
\cos(\theta^{\mathrm{AoD}}_{\mathrm{el}, d_3}) \cos(\theta^{\mathrm{AoD}}_{\mathrm{az}, d_3}),\\
\cos(\theta^{\mathrm{AoD}}_{\mathrm{el}, d_3}) \sin(\theta^{\mathrm{AoD}}_{\mathrm{az}, d_3}),\\
\sin(\theta^{\mathrm{AoD}}_{\mathrm{el}, d_3})
\end{matrix}
\right\rangle.
\end{aligned}
\end{equation}

Using the angle between the vectors \( A_1, A_2, A_3 \), we define:
\begin{equation}
\theta_1 = \cos^{-1} \left( \langle A_1, A_3 \rangle \right), \quad \theta_2 = \cos^{-1} \left( \langle A_2, A_3 \rangle \right)
\end{equation}

\subsubsection*{Distance Estimation}

The UE can estimate the distance to the sensing target using the known distances and angles such as \( d_{\text{1}} \) from TRP to target and \( d_{\text{2}} \) from UE to target.

\begin{equation}
d_{\text{1}} = \frac{d_3^2 + d_4^2 - 2 d_3 d_4 \cos(\theta_2)}{2 d_4 - 2 d_3 \cos(\theta_2)}
\end{equation}
\begin{equation}
d_{\text{2}} = \frac{d_3^2 + d_4^2 - 2 d_3 d_4 \cos(\theta_1)}{2 d_4 - 2 d_3 \cos(\theta_1)}
\end{equation}

\subsubsection*{Target Position Estimation}

Using the AoD angles and TRP coordinates, the 3D position of the target can be estimated as:
\begin{equation}
\begin{aligned}
x_{\text{ST}} &= x_{\text{TRP}} + d_{\text{1}} \cos(\theta^{\text{AoD}}_{\text{el}, d_1}) \cos(\theta^{\text{AoD}}_{\text{az}, d_1}) \\
y_{\text{ST}} &= y_{\text{TRP}} + d_{\text{1}} \cos(\theta^{\text{AoD}}_{\text{el}, d_1}) \sin(\theta^{\text{AoD}}_{\text{az}, d_1}) \\
z_{\text{ST}} &= z_{\text{TRP}} + d_{\text{1}} \sin(\theta^{\text{AoD}}_{\text{el}, d_1})
\end{aligned}
\end{equation}

The joint use of delay and angular measurements significantly increases localization robustness under non-ideal synchronization and multipath conditions.

\section{Simulation Environment}
A 3D simulation environment resembling an indoor factory was used to evaluate the sensing performance. The evaluation uses a system‑level ISAC simulation for an Indoor Factory (InF) hall of \(120\times60\) m with an \(8\) m ceiling. There are 18 TRPs which are uniformly laid out with inter‑site spacing \(D=20\) m and offset \(D/2\) from each wall; and 20 sensing targets are randomly placed on the floor. Propagation includes both LOS and NLOS links for all the paths under the InF setting. 

\begin{figure}[!ht]
    \centering
    \includegraphics[width=0.95\linewidth, height=0.65\linewidth]{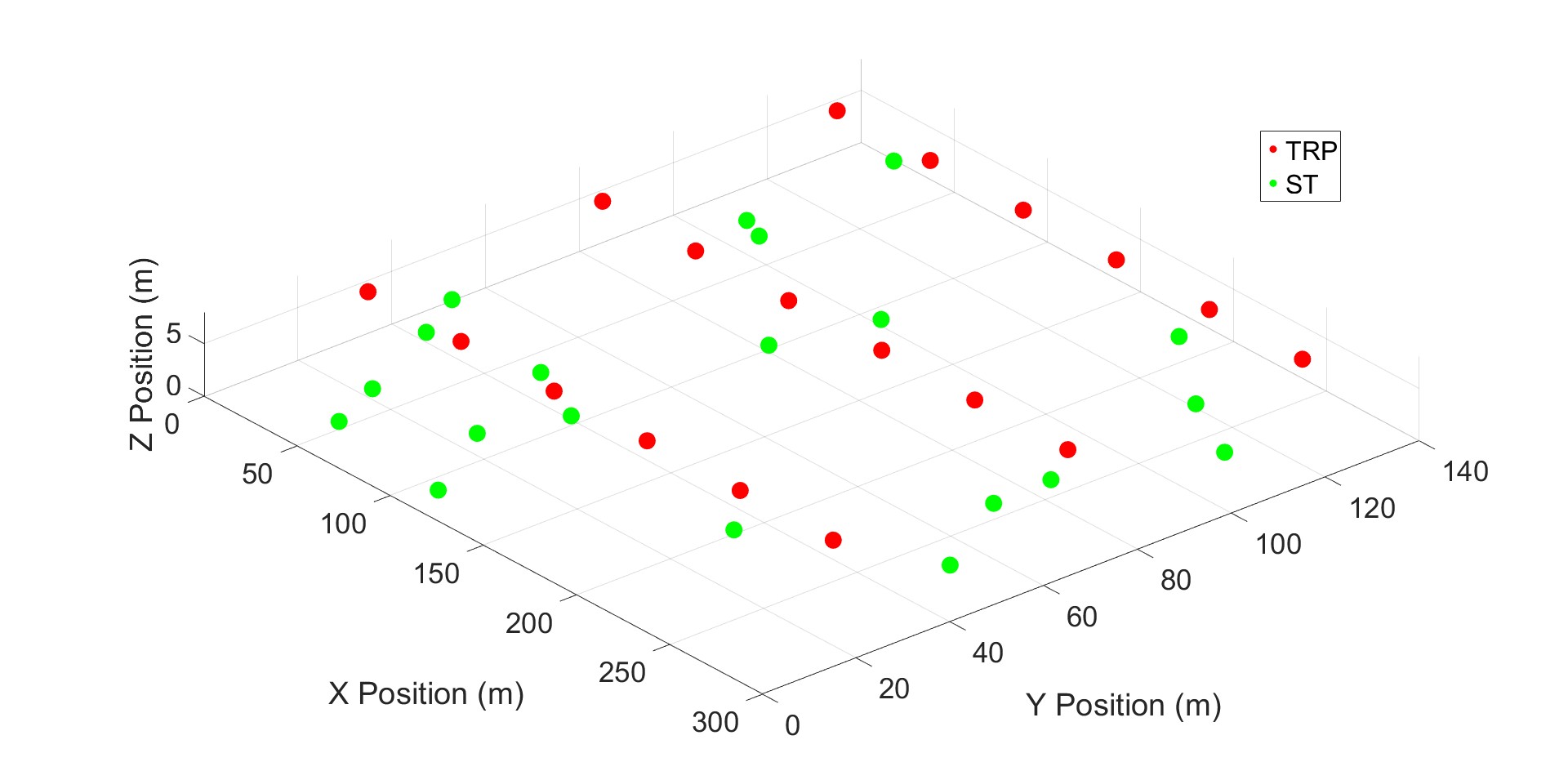}
    \caption{The 3D environment of TRP–TRP bistatic mode.}
    \label{TRP_Env}
\end{figure}

\begin{figure}[!ht]
    \centering
    \includegraphics[width=0.95\linewidth, height=0.65\linewidth]{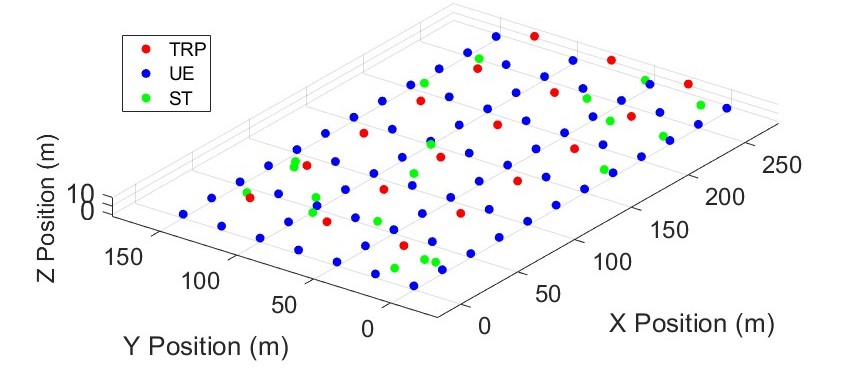}
    \caption{The environment of TRP–UE bistatic and Hybrid mode.}
    \label{TRP_UE}
\end{figure}

Figure~\ref{TRP_Env} illustrates the 3D indoor factory simulation environment configured for the TRP–TRP bistatic sensing mode. Figure~\ref{TRP_UE} presents the simulation environment for the TRP–UE bistatic sensing mode as well as the Hybrid sensing mode. In the TRP–UE mode, the 18 TRPs transmit sensing signals while the 20 UEs receive the reflected echoes.

\section{Target Localization Performance Comparison \\ of TRP–TRP and TRP–UE Bistatic Mode}





\begin{figure}[!ht]
    \centering
    \includegraphics[width=0.95\linewidth, height=0.6\linewidth]{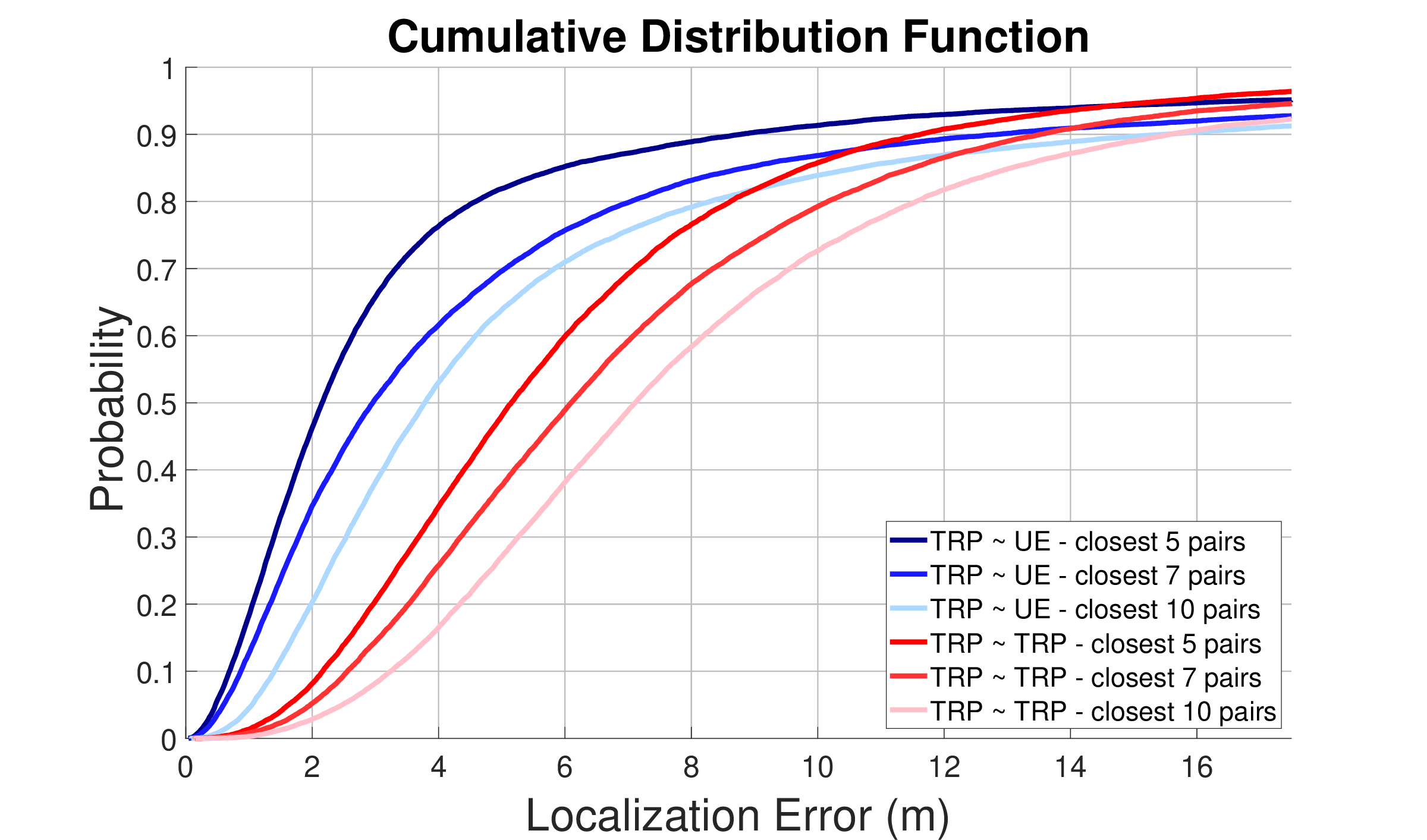}
    \caption{Localization error performed by time-angle based algorithm.}
    \label{fig:cdf_angle}
\end{figure}

\begin{figure}[!ht]
    \centering
    \includegraphics[width=0.95\linewidth, height=0.6\linewidth]{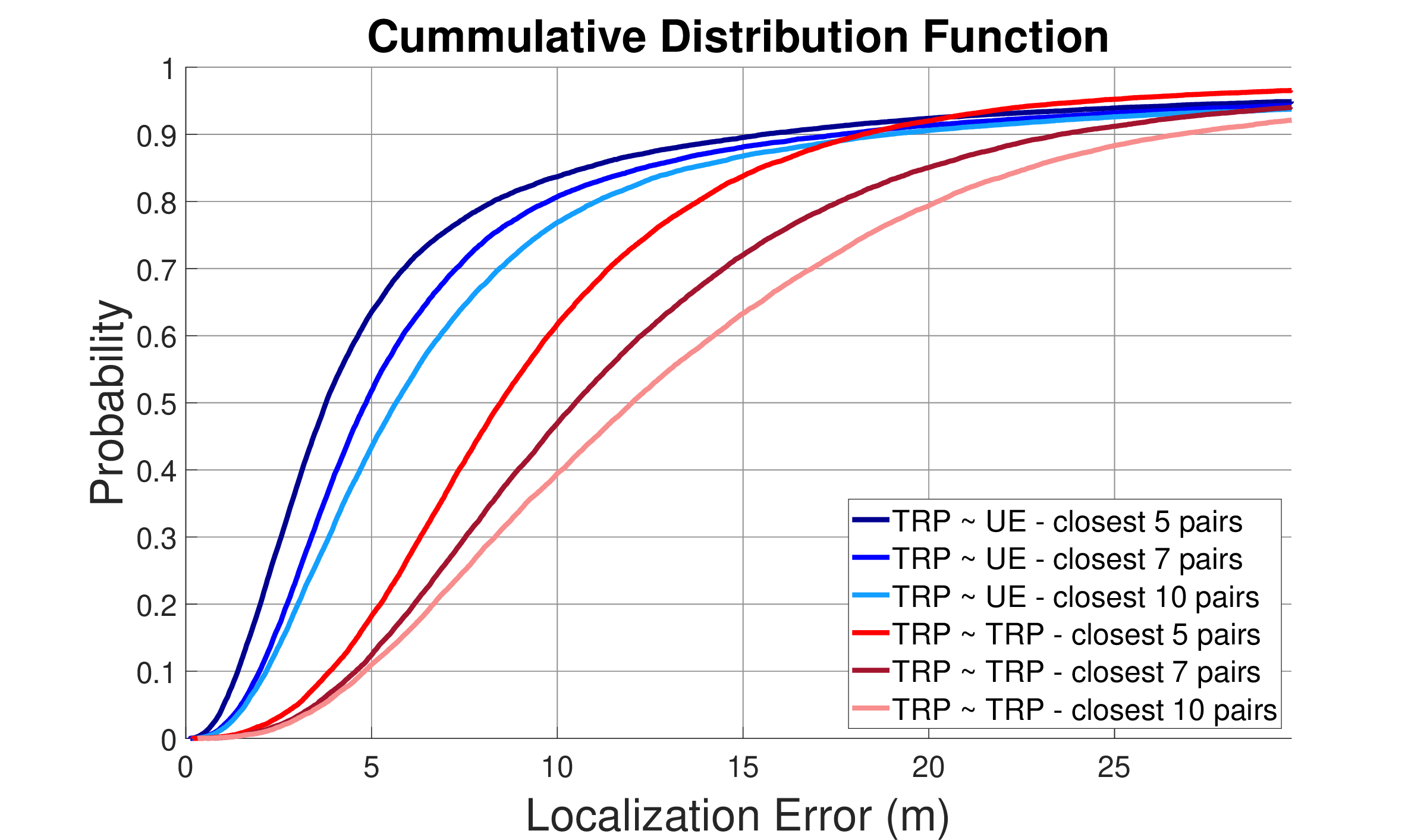}
    \caption{Localization error performed by delay based algorithm.}
    \label{fig:cdf_delay}
\end{figure}

The cumulative distribution functions (CDFs) of sensing accuracy for both bistatic modes are shown in Figure~\ref{fig:cdf_angle} and Figure~\ref{fig:cdf_delay}. The TRP–UE mode outperforms the TRP–TRP mode in terms of accuracy for the closest 5, 7, and 10 sensing pairs. Here, “pairs’’ refers to any transmitter–target–receiver combination used for localization in the given environment. The aggregation of multiple UE measurements contributes to improved sensing performance. This advantage arises from the enhanced spatial diversity and higher probability of LOS conditions when UEs participate in sensing. Overall, the TRP–UE configuration provides better localization precision.

\section{Target Localization Performance of \\ Hybrid Mode Sensing}

\begin{figure}[!ht]
    \centering
    \includegraphics[width=0.95\linewidth, height=0.65\linewidth]{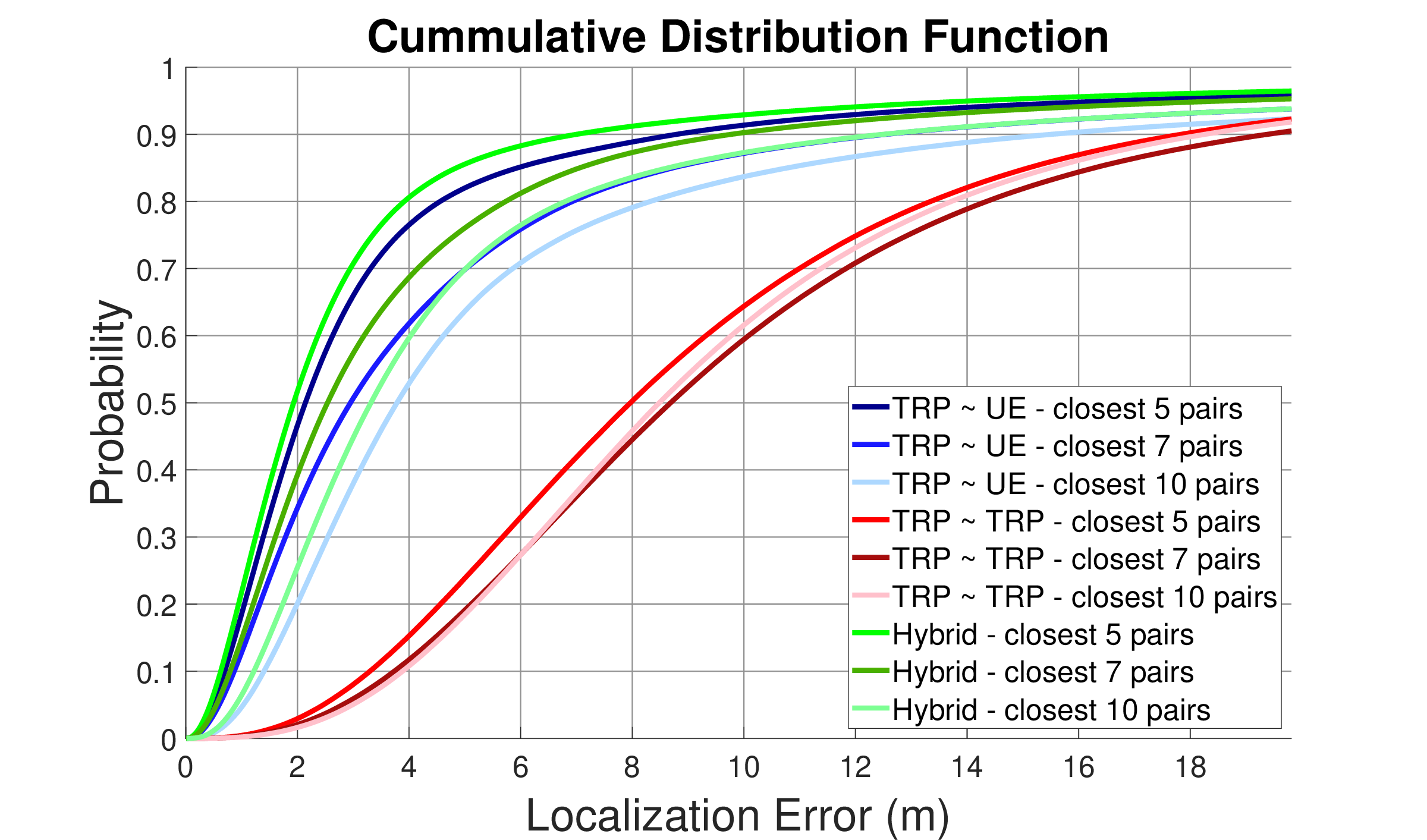}
    \caption{Localization error by time-angle based algorithm}
    \label{fig:cdf_angle_hybrid}
\end{figure}

\begin{figure}[!ht]
    \centering
    \includegraphics[width=0.95\linewidth, height=0.65\linewidth]{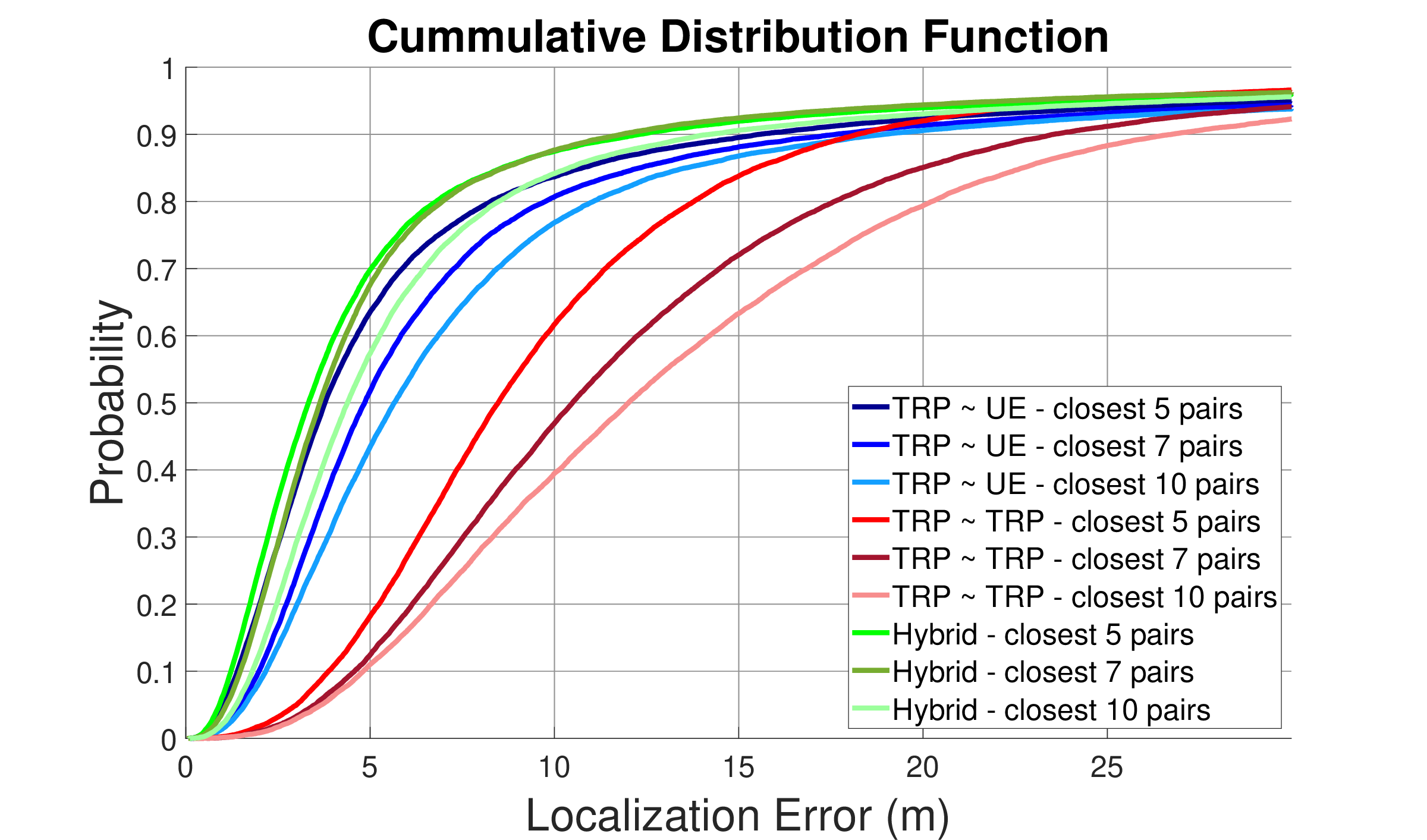}
    \caption{Localization error performed by delay based algorithm.}
    \label{fig:cdf_delay_hbrid}
\end{figure}

The proposed hybrid sensing approach combines bistatic sensing between TRP–UE and TRP–TRP, aiming to exploit the advantages of both modes. By fusing these heterogeneous measurements, the sensing function within the network can significantly enhance localization accuracy and robustness. The localization error CDF across different sensing modes, including the hybrid mode, is presented in Figure~\ref{fig:cdf_angle_hybrid} for the time–angle–based algorithm and in Figure~\ref{fig:cdf_delay_hbrid} for the delay-based algorithm. Hybrid sensing consistently outperforms both TRP–TRP and TRP–UE modes across different pairs. Hybrid sensing exhibits the steepest rise in the CDF, indicating higher localization precision. TRP–UE performs better than TRP–TRP, likely due to a higher likelihood of LOS conditions and more favorable geometry. Aggregating UE-based measurements significantly boosts sensing accuracy, as evidenced by the tighter CDF curves.



Bistatic TRP–UE sensing enhances TRP–TRP sensing by providing a higher probability of LOS conditions and reduced sensing localization error from more diverse angular coverage. Furthermore, aggregating multiple UE-based measurements leads to a substantial increase in overall sensing localization error. As a result, hybrid sensing which has measurement fusion from both TRP–TRP and TRP–UE modes, achieves the best localization performance among all evaluated configurations.

\section{Conclusion}
This study evaluated the localization performance of various integrated sensing and communication (ISAC) sensing modes, including TRP-TRP bistatic, TRP-UE bistatic, and hybrid configurations. The results show that incorporating user equipment (UE) into the sensing process can significantly improve localization accuracy compared with conventional TRP-TRP based sensing. This improvement is mainly due to the additional spatial diversity and more favorable geometric configurations introduced by UE participation. In hybrid sensing, multiple sensing links can be formed between transmission reception points (TRPs), UEs, and the sensing target, allowing the system to collect richer and more complementary measurement information. By combining these measurements, the hybrid configuration can reduce localization ambiguity, mitigate geometric degeneracies, and increase the probability of obtaining at least one high-quality sensing link for each target.

In terms of localization algorithms, time-angle based approaches demonstrated smaller uncertainty regions than delay-only methods. This is because time-angle based localization exploits both timing and angular information, providing more constraints on the target position and improving estimation reliability. The method benefits from initial processing within the sensing environment composed of the TRP, UE, and target, followed by the averaging or fusion of localization results from different sensing pairs. As a result, hybrid sensing combined with time-angle based localization provides a promising and scalable solution for achieving high-accuracy target detection in future 6G networks. Overall, the findings indicate that hybrid mode sensing, particularly when paired with time-angle based localization techniques, can play an important role in enabling precise, robust, and efficient sensing services in next-generation wireless systems.

\section*{Acknowledgment}
This work was funded and supported in part by the MSCA INTEGRATE, and the SNS JU 6G-DISAC Horizon Europe projects under Grant Agreement numbers 101072924 and 101139130, respectively. 

\bibliographystyle{IEEEtran}
\bibliography{references.bib}

@INPROCEEDINGS{10978368,
  author={Uniyal, Smriti and Nguyen, Nhan Thanh and Kumar, Guddu and Di Renzo, Marco and Juntti, Markku},
  booktitle={2025 IEEE Wireless Communications and Networking Conference (WCNC)}, 
  title={Sum Rate and Cramér-Rao Lower Bound Analysis for RIS-Assisted Multiuser Large-Antenna ISAC}, 
  year={2025},
  volume={},
  number={},
  pages={1-6},
  doi={10.1109/WCNC61545.2025.10978368}}

@ARTICLE{10680116,
  author={Paul, Anal and Singh, Keshav and Kaushik, Aryan and Li, Chih-Peng and Dobre, Octavia A. and Di Renzo, Marco and Duong, Trung Q.},
  journal={IEEE Journal on Selected Areas in Communications}, 
  title={Quantum-Enhanced DRL Optimization for DoA Estimation and Task Offloading in ISAC Systems}, 
  year={2025},
  volume={43},
  number={1},
  pages={364-381},
  doi={10.1109/JSAC.2024.3460061}}

@ARTICLE{10833623,
  author={Tishchenko, Anton and Khalily, Mohsen and Shojaeifard, Arman and Burton, Fraser and Björnson, Emil and Renzo, Marco Di and Tafazolli, Rahim},
  journal={IEEE Communications Surveys \& Tutorials}, 
  title={The Emergence of Multi-Functional and Hybrid Reconfigurable Intelligent Surfaces for Integrated Sensing and Communications -A Survey}, 
  year={2025},
  volume={},
  number={},
  pages={1-1},
  doi={10.1109/COMST.2024.3519785}}

@INPROCEEDINGS{10012846,
  author={Erkek, Necati Kagan and Balcı, Emre and Halay, Berkin and Erdemir, Ubeydullah and Görçin, Ali and Çırpan, Hakan Ali},
  booktitle={2022 IEEE 96th Vehicular Technology Conference (VTC2022-Fall)}, 
  title={Measurement-Based Cellular Band Air–to–Ground Channel Modeling for UAVs}, 
  year={2022},
  volume={},
  number={},
  pages={1-6},
  doi={10.1109/VTC2022-Fall57202.2022.10012846}}

@ARTICLE{17564,
  author={Stoica, P. and Nehorai, Arye},
  journal={IEEE Transactions on Acoustics, Speech, and Signal Processing}, 
  title={MUSIC, maximum likelihood, and Cramer-Rao bound}, 
  year={1989},
  volume={37},
  number={5},
  pages={720-741},
  doi={10.1109/29.17564}}

@article{ZhangJSTSP2021,
  author  = {J. A. Zhang and F. Liu and C. Masouros and R. W. Heath and Z. Feng and L. Zheng and A. Petropulu},
  title   = {An Overview of Signal Processing Techniques for Joint Communication and Radar Sensing},
  journal = {IEEE Journal of Selected Topics in Signal Processing},
  year    = {2021},
  volume  = {15},
  number  = {6},
  pages   = {1295--1315}
}

@article{Liu2020JCR,
  author  = {F. Liu and Y. Cui and C. Masouros and J. A. Zhang and L. Hanzo},
  title   = {Integrated Sensing and Communications: Towards Dual-Functional Wireless Networks},
  journal = {IEEE Journal on Selected Areas in Communications},
  year    = {2020},
  volume  = {38},
  number  = {11},
  pages   = {1--17}
}

@article{Liu2022Survey,
  author  = {F. Liu and W. Yuan and Y. Cui and C. Masouros and M. Shafi and L. Hanzo},
  title   = {A Survey on Integrated Sensing and Communication},
  journal = {IEEE Communications Surveys \& Tutorials},
  year    = {2022},
  volume  = {24},
  number  = {2},
  pages   = {778--821}
}

@techreport{3gppTR22837,
  author  = {3GPP},
  title       = {Study on Integrated Sensing and Communication},
  institution = {3GPP SA1},
  number      = {TR 22.837},
  year        = {2024}
}

@article{ZhangVTM2021,
  author  = {J. A. Zhang and M. Rahman and X. Huang and Y. J. Guo and S. Chen and R. W. Heath},
  title   = {Perceptive Mobile Networks: Cellular Networks With Radio Vision},
  journal = {IEEE Vehicular Technology Magazine},
  year    = {2021},
  volume  = {16},
  number  = {2},
  pages   = {20--30}
}

@inproceedings{Bistatic3GPPVTC2024,
  author    = {H. Sun and Y. Zhang and Y. Cui and others},
  title     = {Channel Modeling Framework for Bistatic ISAC under 3GPP Standard},
  booktitle = {Proc. IEEE Vehicular Technology Conference (VTC2024-Spring)},
  year      = {2024}
}

@ARTICLE{10577673,
  author={Zhu, Zhengyu and Li, Zheng and Chu, Zheng and Guan, Yingying and Wu, Qingqing and Xiao, Pei and Renzo, Marco Di and Lee, Inkyu},
  journal={IEEE Internet of Things Journal}, 
  title={Intelligent Reflecting Surface Assisted mmWave Integrated Sensing and Communication Systems}, 
  year={2024},
  volume={11},
  number={18},
  pages={29427-29437},
  doi={10.1109/JIOT.2024.3421319}}

@ARTICLE{10077115,
  author={Masouros, Christos and Zhang, J. Andrew and Liu, Fan and Zheng, Le and Wymeersch, Henk and Di Renzo, Marco},
  journal={IEEE Wireless Communications}, 
  title={Guest Editorial: Integrated Sensing and Communications for 6G}, 
  year={2023},
  volume={30},
  number={1},
  pages={14-15},
  doi={10.1109/MWC.2023.10077115}}

@ARTICLE{8902085,
  author={Koirala, Remun and Denis, Benoît and Uguen, Bernard and Dardari, Davide and Wymeersch, Henk},
  journal={IEEE Access}, 
  title={Localization and Throughput Trade-Off in a Multi-User Multi-Carrier mm-Wave System}, 
  year={2019},
  volume={7},
  number={},
  pages={167099-167112},
  doi={10.1109/ACCESS.2019.2953777}}

@misc{erkek2026communication,
      title={Communication Channel Modelling of Unmanned Aerial Vehicles}, 
      author={Necati Kagan Erkek and Emre Balcı and Berkin Halay and Hakan Ali Çırpan},
      year={2026},
      eprint={2606.16927},
      archivePrefix={arXiv},
      primaryClass={eess.SP},
      url={https://arxiv.org/abs/2606.16927}
}

@misc{erkek2026spaceborne,
      title={Spaceborne SAR Change Detection and Coherence Analysis for Maritime Port Monitoring}, 
      author={Necati Kagan Erkek and Kudret Esmer},
      year={2026},
      eprint={2606.18917},
      archivePrefix={arXiv},
      primaryClass={eess.SP},
      doi={10.48550/arXiv.2606.18917},
      url={https://arxiv.org}
}

@misc{erkek2026reference,
      title={Reference-Based Recursive Least-Squares Mitigation of Real Interference in Stereo Audio Recordings}, 
      author={Necati Kagan Erkek and Y. Ugur Ozcan},
      year={2026},
      eprint={2606.18564},
      archivePrefix={arXiv},
      primaryClass={cs.SD},
      doi={10.48550/arXiv.2606.18564},
      url={https://arxiv.org}
}

@misc{erkek2026joint,
      title={Joint Direction-of-Arrival and Range Estimation for Millimeter-Wave Uniform Linear Array Radar}, 
      author={Necati Kagan Erkek and Zeynep Gul Pehlivanli},
      year={2026},
      eprint={2606.17801},
      archivePrefix={arXiv},
      primaryClass={eess.SP},
      doi={10.48550/arXiv.2606.17801},
      url={https://arxiv.org}
}

@misc{erkek2026pilot,
      title={Pilot-Aided MIMO Channel Identification and Linear Deconvolution in Correlated Gaussian Noise}, 
      author={Necati Kagan Erkek and Y. Ugur Ozcan},
      year={2026},
      eprint={2606.17311},
      archivePrefix={arXiv},
      primaryClass={eess.SP},
      doi={10.48550/arXiv.2606.17311},
      url={https://arxiv.org}
}
\end{document}